\documentclass[pre,floatfix,twocolumn,showpacs]{revtex4}  
\usepackage{graphicx}
\usepackage{amssymb}
\begin{document}

\title{Force networks and elasticity in granular silos}
\author{John F. Wambaugh\footnote{Current address: National Center for Computational Toxicology, US EPA, Research Triangle
Park, NC 27711}}
\affiliation{Department of Physics and Center for Nonlinear and Complex Systems, Duke University, Durham, NC 27708}
\email{wambaugh@phy.duke.edu}
\author{Robert R. Hartley}
\affiliation{Department of Physics and Center for Nonlinear and
  Complex Systems, Duke University, Durham, NC 27708}
\author{Robert P. Behringer}
\affiliation{Department of Physics and Center for Nonlinear and Complex Systems, Duke University, Durham, NC 27708}

\begin{abstract}
We have made experimental observations of the force networks within a
two-dimensional granular silo similar to the classical system of
Janssen.  Models like that of Janssen predict that pressure within a
silo saturates with depth as the result of vertical forces being
redirected to the walls of the silo where they can then be carried by
friction.  By averaging ensembles of experimentally-obtained force
networks in different ways, we compare the observed behavior with
various predictions for granular silos.  We identify several
differences between the mean behavior in our system and that predicted
by Janssen-like models: We find that the redirection parameter
describing how the force network transfers vertical forces to the
walls varies with depth.  We find that changes in the preparation of
the material can cause the pressure within the silo to either saturate
or to continue building with depth.  Most strikingly, we observe a
non-linear response to overloads applied to the top of the material in
the silo.  For larger overloads we observe the previously reported
``giant overshoot" effect where overload pressure decays only after an
initial increase [G. Ovarlez et al., Phys. Rev. E {\bf 67}, 060302(R)
(2003)].  For smaller overloads we find that additional pressure
propagates to great depth.  This effect depends on the particle
stiffness, as given for instance by the Young's modulus, $E$, of the
material from which the particles are made.  Important measures
include $E$, the unscreened hydrostatic pressure, and the applied
load.  These experiments suggest that when the load and the particle
weight are comparable, particle elasticity acts to stabilize the force
network, allowing non-linear network effects to be seen in the mean
behavior.
\end{abstract}


\date{\today}
\maketitle

\section{Introduction}

The physics of granular matter is of great interest both because of
the enormous breadth of physical systems in the granular regime as
well as the lack of a characterization of the granular state in
fundamental terms.  Although there has been some success understanding
energetic granular gases in terms of kinematic models, dense granular
materials continue to pose difficult problems.  We consider dry
granular matter at low humidity where there are no attractive forces
between the constituent particles and the particles do not
significantly interact with the interstitial fluid (air).  Contact
forces dominate dense granular materials in this regime, with each
particle having several persistent contacts.

Coulomb's law of friction is an inequality relating the magnitude of
the friction force, $F_f$ to the normal force, $F_n$, depending
upon the {\em mobilization} of the friction.  If $\mu_s$ is the static
friction coefficient, then the mobilization is given by $F_f/(F_n
\mu_s)$.  For a non-moving contact the mobilization ranges from $-1$
to $1$ and reflects both the direction in which the force of friction
acts to resist motion and how close the forces acting on the contact
are to causing motion in the opposite direction.  Because the force of
friction is dependent upon how the contact arose, the granular state
is determined both by the positions of the particles and the history
of how the particles came to be in these positions
\cite{Clement99a}. Determining a statistical description of these
contacts is at the core of understanding dense granular matter.

\begin{figure}
\centering
\includegraphics[width=2.0in]{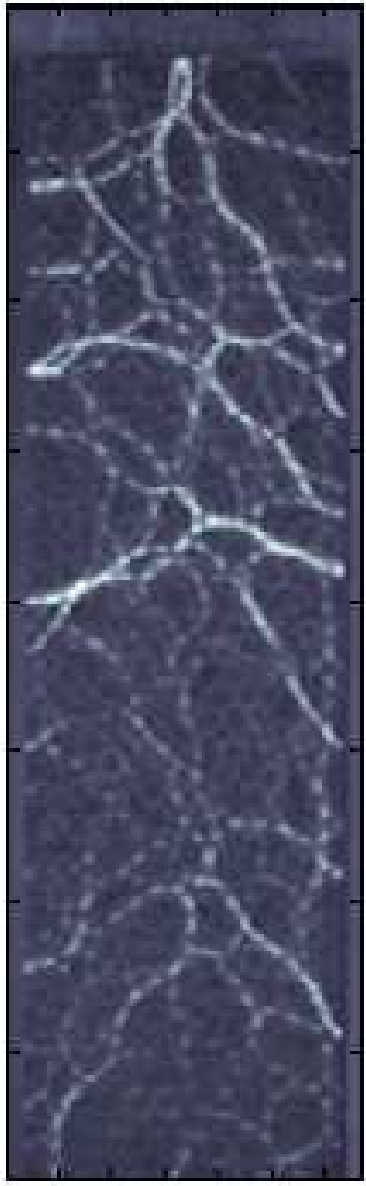}
\caption{A force chain network in a silo filled with photoelastic disks under a $56$ g load}
\label{chains}
\end{figure}
Grain-scale statistical descriptions are further complicated by the
presence of force networks \cite{Clement00,Behringer03}.  An
extraordinary feature of granular contact networks is that the
distribution of forces on the contacts can be strongly
inhomogeneous \cite{Mueth98}.  Certain sequences of contacts carry
forces with magnitude many times the mean for long distances along
``force chains".  As shown in Fig. \ref{chains}, force networks can
develop to support the weight of the particles as well as any
additional applied load.

Despite the difficulties inherent to understanding granular materials,
phenomenological engineering models exist that describe the mean
behavior well enough to allow for the design of granular storage
facilities and handling processes.  These models can provide useful
tools for investigating granular materials.  In particular, Janssen's
model of granular materials in vertical silos is interesting because
it has been thoroughly tested, is realizable in the laboratory, and
ultimately is viewed to be a working qualitative and even quantitative
description of the mean behavior of granular matter
\cite{Janssen,Sperl05}.

\section{The Janssen Model}

Janssen treats the material in a silo as a continuum and considers
horizontal slices, for which the difference of stress, the weight of
the material in the slice, and the force of friction at the wall must
be balanced vertically.  Here we consider a quasi-two-dimensional
silo, such as the one we use in the experiments reported below.
Janssen assumes that the stress does not vary horizontally, allowing
one to write for the case of a two-dimensional silo:
\begin{equation} L\left(\sigma_{yy}\left(y+\Delta\right) -
\sigma_{yy}\left(y\right)\right) =
\rho g L \Delta - 2 \sigma_{xy} \Delta \label{JanssenDif} \end{equation}
where $x$ and $y$ are the horizontal and vertical directions respectively,
$\Delta$ is an infinitesimal change in $y$, $\sigma$ is the stress tensor,
$\rho$ is the density of material (mass per unit area), $g$ is the
acceleration of gravity and $L$ is the width of the silo.  The second term
on the right-hand-side corresponds to the two frictional wall contacts ---
one at each side of the silo.  Coulomb's law of friction gives the shear
stress at the wall in terms of the horizontal stress, $|\sigma_{xy}| \leq
\mu_w|\sigma_{xx}|$ for wall friction $\mu_w$.  Janssen further assumes
that the vertical and horizontal normal stresses are linearly related to
each other to write $\sigma_{xy} = k\mu_w\sigma_{yy}$ where $k$ represents
both the mobilization of interparticle contact friction (with friction
coefficient $\delta$) and the mobilization of friction at the walls.  By
considering the extremes of maximum upward and downward mobilization, it
has been shown that $\frac{1 - \delta}{1 + \delta} < k < \frac{1 +
\delta}{1 - \delta}$ \cite{Nedderman92}.

\begin{figure}
\centering
\includegraphics[width=3.375in]{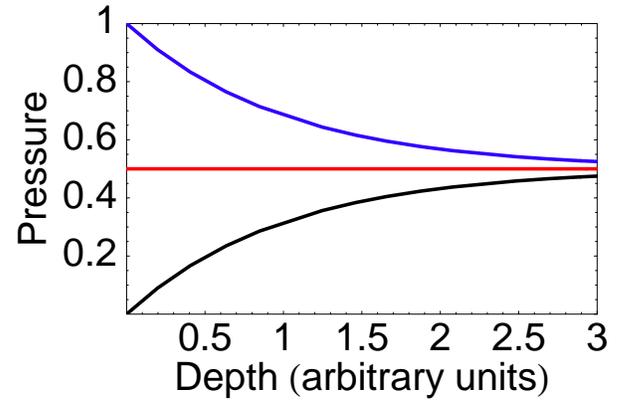}
\caption[Janssen predictions for stress within a silo]{Janssen's model
predicts that stress exponentially approaches a saturation value with
depth in a silo.  Overloads decay exponentially and if an overload
equal to the the saturation stress is placed on the silo the stress is
constant.  Here we plot predictions in a system that saturates at $0.5$ in arbitrary units for no load, a saturation overload ($=0.5$), and an above saturation overload.}
\label{janssenprediction}
\end{figure}
Using Janssen's relation for the shear stress in Eq. \ref{JanssenDif}
allows us to rewrite it as a differential equation for $\sigma_{yy}$:
\begin{equation}
\frac{d\sigma_{yy}}{dy} = -\frac{2 k \mu_w}{L} \sigma_{yy} + \rho g
\end{equation}
If Janssen's parameter $k$ is taken as a constant, then this equation
permits exponential solutions of the form:
\begin{equation}
\sigma_{yy} = \Phi_0 e^{-\frac{y}{\lambda}} + \frac{\rho g L}{2 k \mu_w}\left(1 - e^{-\frac{y}{\lambda}}\right)
\end{equation}
where $\Phi_0$ is any load applied to the top of the system and
\begin{equation}
\lambda \equiv \frac{L}{2 k \mu_w}
\label{eq:lambda}
\end{equation}
 sets a length-scale for the evolution of stress with depth.  As
indicated by Fig.~\ref{janssenprediction}, the Janssen solution
predicts that any overload will decay exponentially, and that the
stress within the material will saturate with depth.  This {\em
screening} effect results from friction at the walls acting to carry
the weight of the material and any overload.  Although we have
presented the two-dimensional version here, the Janssen model is
qualitatively similar in three dimensions.

Janssen's prediction of stress saturating with depth has been
validated qualitatively in numerous three-dimensional experiments that
examine the apparent weight of a confined granular pile.  Though these
experiments in effect only measured the forces at the bottom of the
pile, by repeatedly varying the amount of material --- and thus the
height of the pile --- a mean description of stress with depth was
obtained
\cite{Clement97,Clement99b,Clement00,Clement03,Hulin03,Hulin04}.
Large scale simulations of three-dimensional systems have also
observed stress saturation \cite{Landry03a}.  In quasi-two-dimensional
experimental systems that are similar to the analysis above, the
saturation length scale has been found to scale with system size as
predicted, $\lambda \propto L$ \cite{Peralta-Fabi97}.

While qualitatively accurate, Janssen's predictions have been shown to
underestimate the actual value of saturated stress in a silo
\cite{Clement99b}.  Careful analysis of stress profiles in experiments
and simulations have shown that for small depths the profile is more
linear than exponential \cite{deGennes97,Clement99b}.  These
deviations from the Janssen model are not surprising, since key
assumptions of the analysis can be shown to be incorrect.

For the Janssen model above we assumed that the redirection parameter
$k$ was constant.  For a cohesionless granular material, however, the
ratio of the horizontal to vertical stress is constant only if the
horizontal and vertical directions are the frame in which the stress
tensor is diagonal \cite{Nedderman92}.  This is inconsistent, since
Janssen assumes that there is a non-zero $\sigma_{xy}$ at the walls.
If $\sigma_{xx}$ and $\sigma_{yy}$ are not the principal stresses,
then we expect $k \sim \sigma_{xx} / \sigma_{yy}$ to vary horizontally
and with depth.

Several factors contribute to variation in the redirection parameter
with depth.  In particular, it is difficult to achieve a uniform
mobilization of friction at the walls.  Experiments and
simulations have shown that agreement with Janssen predictions can be
obtained if the granular material is slowly lowered relative to
the walls to make uniform the direction and extent of mobilization for
all wall contacts \cite{Pitman98,Hulin03,Landry03c}.  The global
packing fraction of the material has also been shown to alter the
redirection parameter and it is reasonable to expect that local
variations in packing fraction would produce variations with depth
\cite{Clement03}.  Finally, experiments studying the stress within a
silo of deformable beans, instead of the more rigid glass or metal
balls typically used, have found that the redirection parameter varies
with depth due to the elasticity of the particles \cite{Atewologun91}.

More elaborate analysis of the Janssen model using stochastic
differential equations to account for randomness in the mobilization
of friction has shown that the Janssen model only correctly predicts
the stress on the axis of the silo \cite{Pitman98}.  In general, the
vertical stress at a given depth has been found to vary horizontally,
further complicating the force balance analysis
\cite{Nedderman92,Rusinek03}.  In three dimensions, attempts have been
made to account for radial stress balance by using vertical force
balance on concentric rings about the axis with the general result
that force saturation due to screening by the walls is still expected,
though the approach to the saturation value may not be exponential
\cite{Lvin70}.

Although screening-induced saturation has been validated, the behavior
of overloads can be strikingly different from predictions.
Instead of a monotonic, exponential decay to the saturated stress
supported by the walls, a non-monotonic, ``giant overshoot" has been
observed experimentally \cite{Clement03,Clement05}.  Even an overload
equal to the saturated stress deep in the silo produces a local rise
in the stress instead of the flat stress profile predicted by the
Janssen model.  This can in part be explained by local changes in the
mobilization of friction and packing fraction of granular matter
induced by placing an overload on the material.

To account for the overshoot effect and better fit experimentally
observed stress profiles, a piece-wise model has been proposed that
describes stress in a silo as hydrostatic - increasing linearly with
depth - until a tunable depth where the stress then becomes
exponential \cite{deGennes97,Clement99b}.  This piece-wise behavior
arises out of the analysis for horizontally varying vertical stress
and indicates the depth to which the stress at the top boundary
penetrates \cite{Lvin70}.  Fitting experimental results to the
piece-wise pressure profile corrects experiments that find unphysical
values of the redirection parameter when the basic Janssen model is
used \cite{Clement99b,Clement03}.

In some experiments, a threshold for overloads has been observed that,
when passed, causes the redirection parameter $k$ to increase
\cite{Rusinek03}.  When the presence of force networks within granular
silos is considered, the mechanism accounting for different overload
profiles becomes clearer.  Granular matter has been described as being
in a ``fragile" state.  Along the directions in which there are
pre-existing force chains, large loads can be supported.  In other
directions, only slight deformation of the interparticle contacts are
allowed before the material must rearrange, forming new force chains
\cite{Claudin98}.  To understand how a granular silo responds to
overloads, we need to consider how the force network facilitates the
redirection of stress to the walls, which is the goal of the
present experiments.

Since the Janssen model is a continuum description that at least
moderately successfully predicts stress saturation with depth, it is
interesting to examine in what ways it gives an appropriate mean
description of the forces within a granular material.  It is equally
interesting to determine how and why it fails
\cite{Clement99a,Clement01d}.  For instance, various models of
granular force propagation give different predictions of mean silo
behavior \cite{Coppersmith96,Socolar97,Clement00}.  Further
experimental analysis of the silo system can serve to test these and
future models of granular matter.

In our experiments, we use photoelastic particles to investigate how
the preparation and properties of the granular material and the
network of contact forces interact to redirect stress to the walls and
determine the development of the stress profile with depth.  We
conduct experiments in a quasi-two-dimensional system that is
described by the Janssen analysis above and has been shown to be
similar to three dimensional systems in experiments and simulations
\cite{Peralta-Fabi97,Baxter97,Landry03b}.  Similar
quasi-two-dimensional experiments have used photoelasticity in the
past to examine the propagation of stress but did not test the Janssen
model \cite{Baxter97,Clement01a,Behringer03}.  Rather than measure the
apparent weight at the bottom of a three-dimensional pile, these
photoelastic techniques allow us to examine the actual force network
responsible for Janssen screening.  We investigate the role of
particle elasticity and force networks within silos by examining the
mean response to overloads, by characterizing the properties of the
force network, and by varying the elasticity of the particles.

\section{Methodology}

\begin{figure}
\centering
\includegraphics[width=3.375in,clip=]{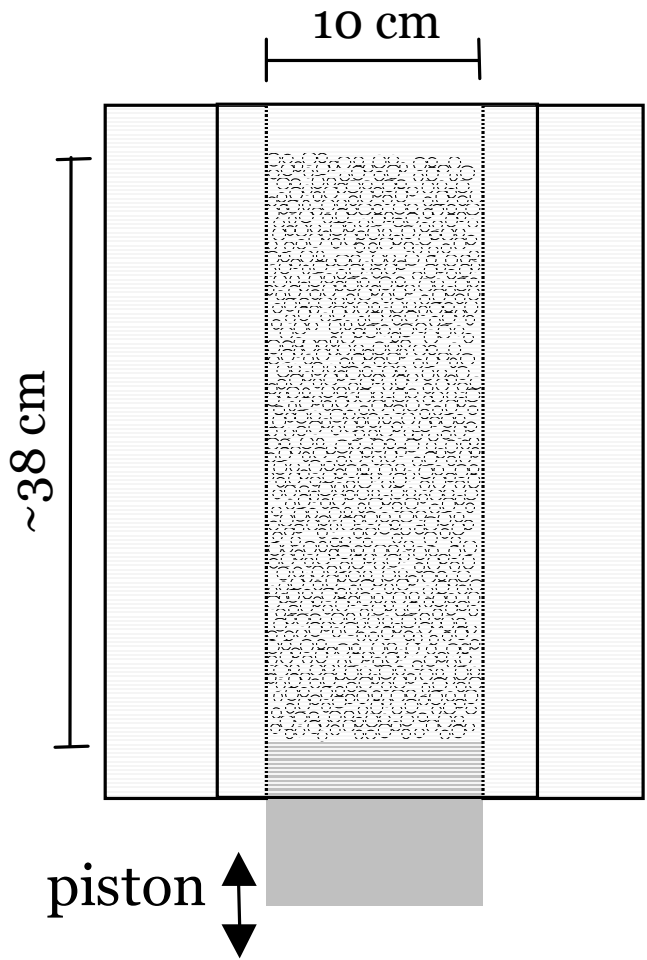}
\caption[Granular silo experimental apparatus]{We examine photoelastic
disks constrained to a quasi-two dimensional geometry in a vertical
`silo' made from two Plexiglas sheets separated by slightly more than
a disk thickness.  The bi-disperse disks are confined by vertical
aluminum side walls to a channel $L = 10$ cm wide
and rest upon a piston that can be used to control the mobilization of
friction.
The aspect ratio of the pile is typically 4:1.}
\label{silopapersetup}
\end{figure}
We study a quasi-two dimensional arrangement of photoelastic disks
constrained vertically by two aluminum walls and on the faces by two
transparent sheets to form a channel $L = 10$ cm wide, as depicted in
Fig.~\ref{silopapersetup}.  To prevent ordering, we use a bi-disperse
mixture of disks, with the larger disks approximately 20\% larger.  The silo is constructed so that there are movable pistons at the top and bottom of the silo.  The pile
rests upon the bottom piston, while loads can be applied to the top
piston.  A $10 g$ load applied to the top of the pile corresponds to a
$\sim 1000$ dyne/cm stress at the top.  Both pistons can be pinned to
prevent motion or unintended loading.

We measure the local disk-scale pressure, which we refer to as
``stress'' on our photoelastic disks using a technique similar to
previous granular research from our group \cite{Behringer03}.  The
bi-refringent disks display bright fringes in response to applied
stress when illuminated between crossed polarizers.  The density of
fringes within a given disk is proportional to the stress within the
disk.  We determine the stress on a given disk by first calculating
the gradient of an image of fringes.  We take the square of the
gradient averaged over multiple directions to be proportional to the
number of fringes.  We use an experimentally-obtained calibration to
convert the squared gradient of the intensity into an equivalent load
on a monolayer of disks and hence a single disk.

We performed experiments with two different silo/camera arrangements.
In the first silo we use approximately $650$ disks --- one quarter
with diameter $0.9$ cm and the rest $0.75$ cm --- resting upon a
piston that is affixed to a stepper motor.  A $640\times 480$ digital
video camera is used to image the column at a rate of 30
frames/s while the piston supporting the pile is slowly lowered at a
rate of hundredths of a grain diameter per second.  We examine the
sequence of images to find the first frame where an abrupt
rearrangement of the force network is observed and take the frame
before that as the moment that the friction of the grains at the walls
is maximally mobilized upwards.  We perform our analysis on this
maximally mobilized frame.  To prepare for each observation, a new
force network is created by quickly forcing the lower piston upwards
and then letting the entire pile rain downwards as the piston falls.
Simulations have shown that granular piles created by `raining'
particles from above are independent of the rate at which grains are
added \cite{Landry03a}.  To apply an overload we gently lower the
upper piston onto the top of the pile before the pile is lowered by
the stepper motor.  The top layers of grains rearrange slightly as the
overload is applied but the remainder of the particles maintain their
relative positions.

In the second silo arrangement, approximately $1200$ slightly smaller
disks are used (an approximately 3:1 mixture of $0.6$ cm and $0.75$ cm
disks).  The lower piston rests upon a base that can be manually
lowered a small distance.  Without a fixed connection to a stepper
motor, we are able to create new force networks by physically lifting
the entire silo and flipping end-over-end.  After the pile is lowered
slightly to mobilize friction more uniformly, a high-resolution
($3264\times2468$) digital still camera is used to take two pictures,
one without polarization that can be used to find particle centers and
one with polarization for obtaining force from photoelasticity.  We
can use this information to characterize the contact and force
networks within the material.  To investigate the influence of
particle elasticity on these networks, we use disks made from two
photoelastic polymers of different stiffness (Young's modulus $E = 4$
MPa for softer particles, $E = 210$ MPa for harder particles).  Both soft and hard disks are $0.635$ cm thick, and the material from which they are made has a density of
$1.2g/cm^3$.

For all figures, the standard error indicated by bars for mean
quantities is determined using a `bootstrap' algorithm where the mean
of the quantity is calculated for each of $200$ randomized data sets
constructed by selecting elements of the original data set at random
with replacement \cite{Efron91}.  The standard deviation of these
means is taken to be the standard error.

\section{Results}

\subsection{Fixed Silo Method--Softer Particles}

\begin{figure}
\centering
\includegraphics[width=3.375in]{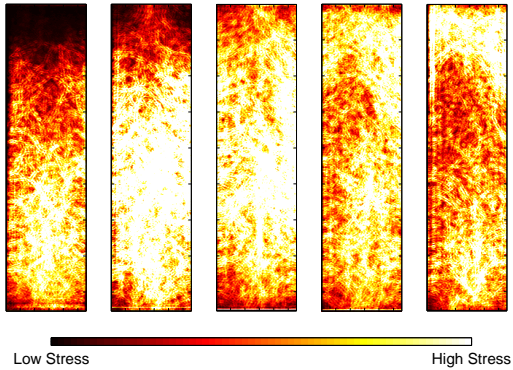}
\caption[Merged images of stress response for various overloads]{(Color Online) These
composite images, generated by merging sets of seventy to one hundred
gradient squared images of force networks, display the onset of
Janssen screening as the overload is increased from $0$ to $106$ g in
approximately $25$ g increments.  The same intensity scale is used for all images, indicating greater pressure deep in the silo for smaller loads.}
\label{mergedpics}
\end{figure}
\begin{figure}
\centering
\includegraphics[width=3.375in]{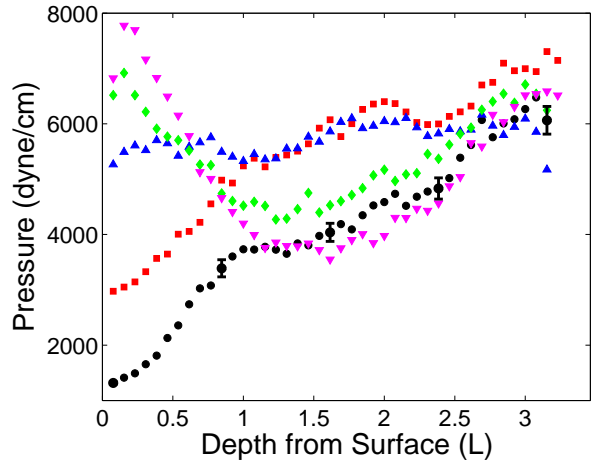}
\caption[Pressure profile with depth in a granular silo]{By binning
the images of force networks horizontally, averaging sets of seventy
to one hundred images and then using a calibration to convert
intensity gradient into pressure, we can generate pressure profiles.
The dashed line indicates the unloaded case, while the increasing
solid lines indicate overloads applied at the top increasing in
increments of roughly $25$ g: no load ($\bullet$), $31$ g
($\blacksquare$), $56$ g ($\blacktriangle$), $81$ g ($\blacklozenge$)
and $106$ g ($\blacktriangledown$).  Error bars indicate the standard
deviation of means calculated for an ensemble of subsets of the data.
Depth is given in terms of the width of the system $L = 10$ cm
$\approx 13$ smaller disk diameters.}
\label{pressure}
\end{figure}
We first used the fixed silo and stepper-motor arrangement to observe
$\sim100$ unloaded force networks and $\sim75$ networks each for four
different overloads.  Figure \ref{mergedpics} shows the merged squared
gradient images of these force networks, roughly depicting mean response.
If we bin the responses across the width of the silo into horizontal
slices the height of one particle, we can obtain a pressure profile with
depth as in Fig.~\ref{pressure}.

Considering first the case of the silo without an overload, we see
distinctly non-Janssen behavior.  Although the pressure initially
seems to saturate with moderate depth, the pressure starts increasing
again toward the bottom of the silo.  This increase is also observed
for all overloads applied to the top of the pile.  This effect is at
least partly due to the way that the preparation history affects the
unloaded state.  As seen in the left-most image of Figure
\ref{mergedpics}, the mean stress is not constant across the layer at
a given height, particularly near the bottom.  This effect is
presumably associated with the fact that when the piston pushes
upwards, not all the particles are easily thrown upward, or that
there is wall drag as the particles fall which affects their packing.

We next examined the response to four overloads in increments of roughly $25$ g
(i.e. a two-dimensional stress of roughly $2500$ dyne/cm).  As shown in
Fig.~\ref{pressure}, for the two largest overloads a non-monotonic
overshoot effect is observed.  This behavior is similar to previous
experiments \cite{Clement03,Clement05} -- after an initial increase in
pressure at the top of the silo, the largest overloads are screened by
arching and the pressure converges with the unloaded pressure profile
with depth.

\begin{figure}
\centering
\includegraphics[width=3.375in]{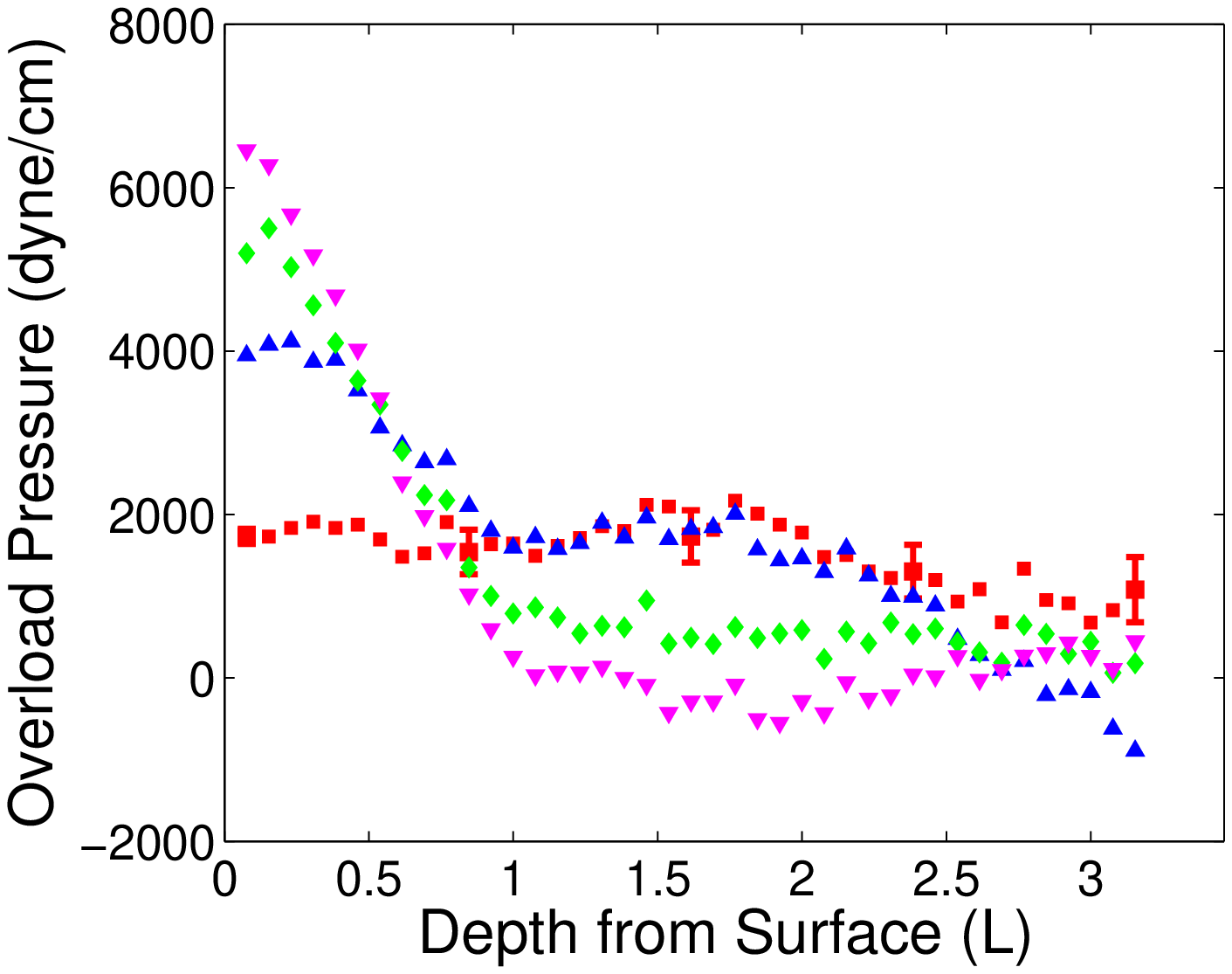}
\caption[Overload profile with depth]{Plotting the difference between the
pressure profile resulting from an overload and the unloaded profile shows
how the overload is distributed.  Four different overloads,
$31$ g ($\blacksquare$), $56$ g ($\blacktriangle$), $81$ g
($\blacklozenge$) and $106$ g ($\blacktriangledown$), were applied across
the top of the system.
Smaller loads seem to penetrate deeper into the pile, while larger loads
seem to show the previously observed overshoot effect \cite{Clement03}
followed by the expected decay.}
\label{overload}
\end{figure}
The behavior of the two smaller overloads is strikingly different from
expected and previously-observed behavior.  The additional pressure of
small overloads persists with depth and is not screened by wall
friction.  The difference between large and small overloads is clear
in Fig.~\ref{overload}, where the unloaded pressure profile has been
subtracted from the four loaded pressure profiles.  Deep within the
pile, the additional pressure of two larger overloads decays to zero,
while the pressure from the smaller two loads persists.

\begin{figure}
\centering
\includegraphics[width=3.375in]{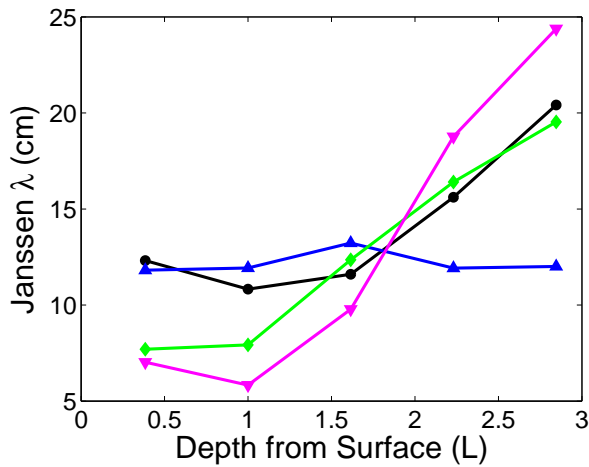}
\caption[Janssen length-scale with depth]{The Janssen length-scale
$\lambda$ can be calculated numerically using the density of the
granular material and the pressure profiles for no load ($\bullet$),
$56$ g ($\blacktriangle$), $81$ g
($\blacklozenge$) and $106$ g ($\blacktriangledown$).  We observe that
$\lambda$ varies as a function of depth instead of remaining constant
as in the Janssen model.  Since the calculated length-scale varies
with the inverse of the derivative of pressure, variations in the
measured pressure profile can be greatly amplified, as with the for
$31$ g results at shallow depth.}
\label{lambdaplot}
\end{figure}

Given the non-monotonic, and sometimes non-saturating behavior of the
observed pressure profiles, it is not surprising that the Janssen
length-scale $\lambda = \frac{L}{2 k \mu_w}$ varies with depth, as indicated in Fig.
\ref{lambdaplot}.  Using our data, we estimate $\lambda$ numerically
as:
\begin{equation}
\lambda(z) = \frac{P(z)}{\rho_{ps}\phi g - \frac{dP(z)}{dz}}
\label{eq:local-lambda}
\end{equation}
where $\rho_{ps}$ is the density of the photoelastic material, $\phi$
is the packing fraction in the silo, $P(z)$ is the horizontally-binned
pressure and $\frac{dP(z)}{dz}$ is estimated numerically using local
quadratic fits to $P(z)$.  Since $\lambda$ varies as the inverse of
the redirection parameter $k$, this result indicates that the
assumption that $k$ is constant with depth within our system is
incorrect.  Though the scatter in the pressure profile for $31$ g
introduces substantial noise into the calculation of $\lambda$,
hence this data is omitted, we see that the variation of
$\lambda$ with depth is roughly similar for both unloaded and loaded
silos.  This indicates that the previously-observed load-induced
variation in $k$ is not responsible for variation in load propagation
\cite{Rusinek03}.

\begin{figure}
\centering
\includegraphics[width=3.375in]{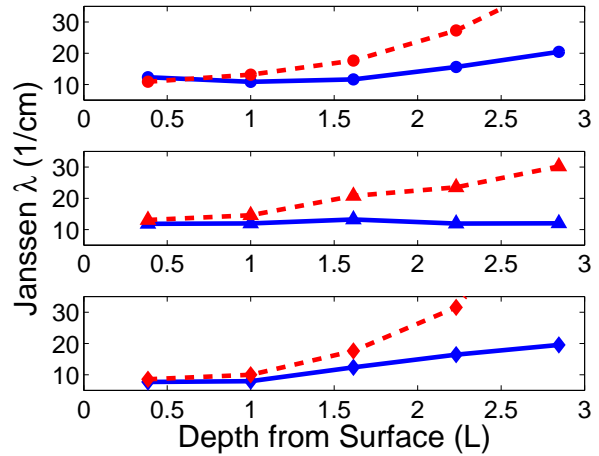}
\caption[Janssen length-scale as a function of mobilization
procedure]{The Janssen screening length-scale $\lambda$ as a function
of depth for, from top to bottom, no load ($\bullet$), a $31$ g load
($\blacktriangle$)and an $81$ g ($\blacklozenge$) load for both the
initial ({\bf dashed}) force network and the maximally mobilized ({\bf
solid}) force network.}
\label{mobilization}
\end{figure}

Though Janssen's $k$ varies with depth, we can still investigate if
$k(z)$ behaves as a depth-dependent redirection parameter might be
expected to behave.  In Fig.~\ref{mobilization}, we calculate
$\lambda$ as a function of depth for the ensemble of unloaded
networks, and the ensembles for overloads just above and below the
transition from deep-propagation to screening.  We compare the network
in the first image of our lowering sequence, corresponding to the
initial network before the pile has been lowered to change
mobilization, to the network in the subsequent image where we assume
friction is maximally mobilized.  Since $\lambda$ varies as the
inverse of the mobilization $k$, the observed decrease in
$\lambda$ after lowering the pile indicates that we have increased
the mobilization.  The mobilization does not change significantly at
the top of the pile, indicating that our lowering technique does not
uniformly effect the pile.  The difference between initial and
mobilized frames is similar for both small, deep-propagating overloads
and large, decaying overloads.

\begin{figure}
\centering
\includegraphics[width=3.375in]{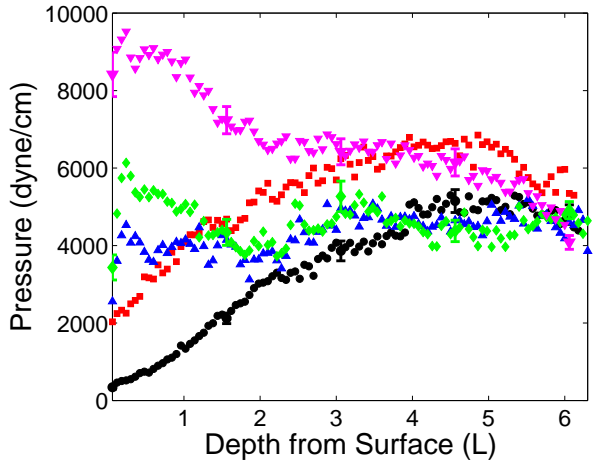}
\caption[Pressure profile with depth using different network creature
procedure]{By changing our procedure for creating new force networks
we obtained pressure profiles that indicate pressure saturation with
depth for an unloaded silo ($\bullet$) and overloads of $20$ g
($\blacksquare$), $40$ g ($\blacktriangle$), $60$ g ($\blacklozenge$)
and $80$ g ($\blacktriangledown$).  Despite the change in pile
creation procedure the overshoot effect is still observed for large
loads and small loads are still observed to penetrate to a
considerable depth.}
\label{newsoftpressure}
\end{figure}
We note four important differences between the pressure profiles in
Fig.~\ref{pressure} and those predicted by Janssen-like models,
e.g. Fig.~\ref{janssenprediction}: $\lambda$, and hence $k$, is not
constant; pressure from larger overloads first increases with depth
before decaying; overall pressure does not seem to saturate with
depth; and most strikingly, pressure from small overloads does not
decay.  The first two differences have been observed previously and
are not surprising, given the previously discussed drawbacks of
Janssen's analysis \cite{Atewologun91,Clement03,Clement05}.  To better
understand the lack of overall pressure saturation, we conducted
experiments in our second apparatus that allowed the particles to be
rained from above, rather than pushed from the bottom.  As shown
in Fig.~\ref{newsoftpressure} we found that the second preparation
procedure did create pressure profiles that saturate with depth.  The
largest overload is observed to overshoot before decaying and the
smallest load propagates to great depth.

\begin{figure}
\centering
\includegraphics[width=3.375in]{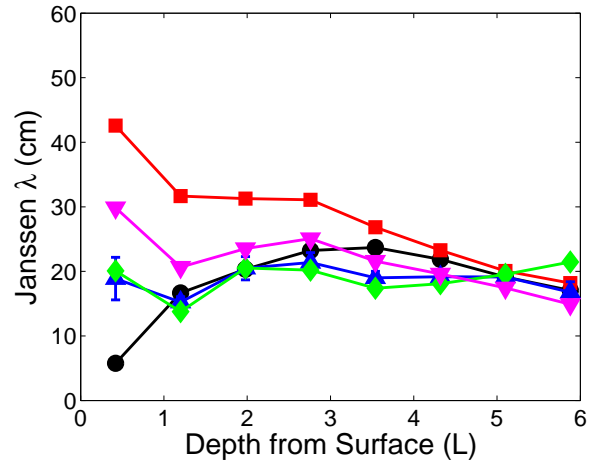}
\caption[Janssen length-scale with new procedure]{The numerically
evaluated Janssen screening length-scale for the pressure profiles for an
unloaded silo ($\bullet$) and overloads of $20$ g ($\blacksquare$), $40$ g
($\blacktriangle$), $60$ g ($\blacklozenge$) and $80$ g
($\blacktriangledown$) shows convergence with depth to a nearly constant
value of $\lambda$.}
\label{newlambdaplot}
\end{figure}
Numerically determining the Janssen screening-length $\lambda$ for the
saturating pressure profiles via Eq.~\ref{eq:local-lambda} indicates
that after an initial transient, the length-scale actually converges
to a constant value of roughly $20$ cm with depth, as in
Fig.~\ref{newlambdaplot}.  This implies a nominal value of $\mu k
= L/2\lambda = 0.25$.  Interestingly, the $\lambda(z)$ for the
moderate overloads are the flattest, while the screening length
converges from a larger value for the larger overloads and from a
smaller value for the silo with no overload and the smaller
deep-penetrating $20$ g overload.  The non-constant value for the
unloaded case may be due to several causes.  One may be an
underestimation of the small forces at the top of our pile when using
our gradient method to calibrate photoelastic fringes.  Alternatively,
the variation at the top of the pile may be explained as a combination
of both preparation --- the act of placing an overload at the top ---
and non-uniform response to lowering the pile with depth.  These two
factors may also account for the giant overshoot effect seen for the
large overload.

\begin{figure}
\centering
\includegraphics[width=3.375in]{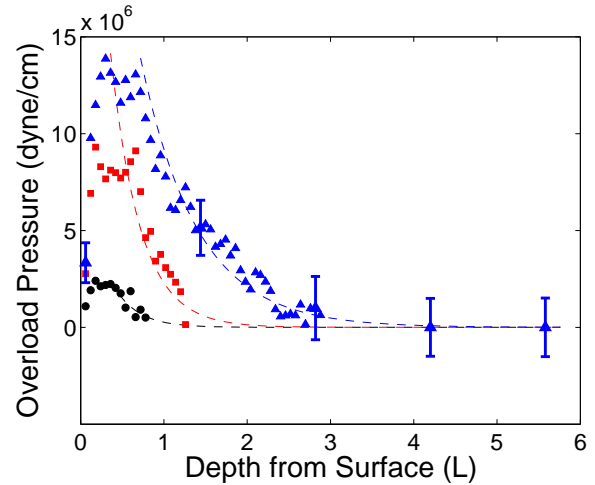}
\caption[Overload profile for harder particles]{For hard ($E = 210$ MPa) photoelastic disks we only see the overload profile due to the
relative insensitivity of the harder material to stress.  We examined
overloads of $1837$ g ($\bullet$), $2267$ g ($\blacksquare$) and $3402$ g
($\blacktriangle$).  All seem well-described by exponential fits ({\bf dashed lines}).}
\label{overloadhard}
\end{figure}

\subsection{Silo Flipping Method--Softer and Harder Particles}

We then used the silo flipping preparation to examine both the effect
of preparation and of changing the elasticity of the particles, from
Young's modulus $E = 4$ MPa to $E = 210$ MPa.  For the harder
photoelastic material, the sensitivity to applied stress was lower,
with the result that the stress due to gravity was not visible even at
great depth in the silo.  In this case, any pressure profile is
effectively an overload decay profile, since there is no unloaded
profile to subtract.  As shown in Fig.~\ref{overloadhard}, the
additional pressure due to the overloads decays with depth after
overshooting for all three cases.  For all three overloads, the mean
pressure eventually drops below the range of observable sensitivity,
though large fluctuations about the mean were still present.  All
three cases can be fit with exponentials if the initial overshoot is
neglected.  Since the non-decaying, deep-penetrating overloads for the
softer particles were close to the saturation pressure, it is not
possible to determine whether or not such effects are present in the
harder particles using our method.

\begin{figure}
\centering
\includegraphics[width=3.375in]{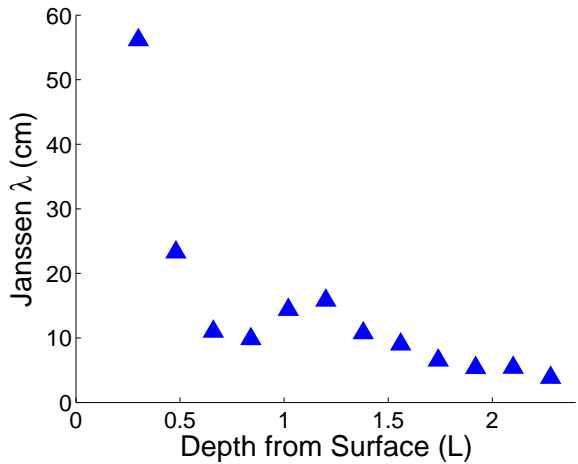}
\caption[Janssen length-scale for harder particles]{Due to the
insensitivity of the harder photoelastic particles, determining the
{\bf local} Janssen screening length-scale $\lambda$ numerically is
only possible for the largest $3402$ g ($\blacktriangle$) overload.
The observed screening length is roughly half as long as for the soft
($E = 4$ MPa) particles.}
\label{hardlambdaplot}
\end{figure}
Determining the Janssen screening length-scale $\lambda$ numerically
for the hard particle overloads was problematic since the pressure
profiles eventually decayed below the sensitivity of the hard
photoelastic material.  In Fig.~\ref{hardlambdaplot}, we show the
calculated $\lambda(z)$ for the largest, $3402$ g overload in the top
third of the silo.  What we see is qualitatively similar to the
behavior of the largest overload in Fig.~\ref{newlambdaplot} ---
$\lambda$ approaches a constant value from above.  Surprisingly,
although the system geometry is the same, the screening length is
roughly half of what was found for the soft ($E = 4$ MPa) particles
used to obtain the first two pressure profiles (Fig.~\ref{pressure}
and Fig.~\ref{newsoftpressure}).

The two sets of experiments nominally contrast soft and hard
particles.  However, there is another aspect which may be at least as
important as granular elasticity.  We note that the ratio of the
applied loads to the Young's modulus, $E$, are comparable for the hard
and the soft material.  For the maximum load on the softer particles,
$\sim 100g$, this ratio is $\sim 1.6 \times 10^{-4}$ and for the
maximum load on the harder particles, $\sim 3400 g$, this ratio is
$\sim 1.0 \times 10^{-4}$.  Thus, in appropriately scaled terms the
loads are comparable in both cases, and hence on this basis,
elasticity cannot explain the difference in the experimental
observations.  However, the ratio of the applied load to the weight of
the material is significantly different in the two cases, namely about
0.5 for the soft particles, and 17 for the hard particles.  Thus, we
conclude that for the softer particles, the loads are not large enough
to overcome the effects caused by gravitational loading, which
necessarily reflects the preparation history.

\begin{figure}
\centering
\includegraphics[width=3.375in]{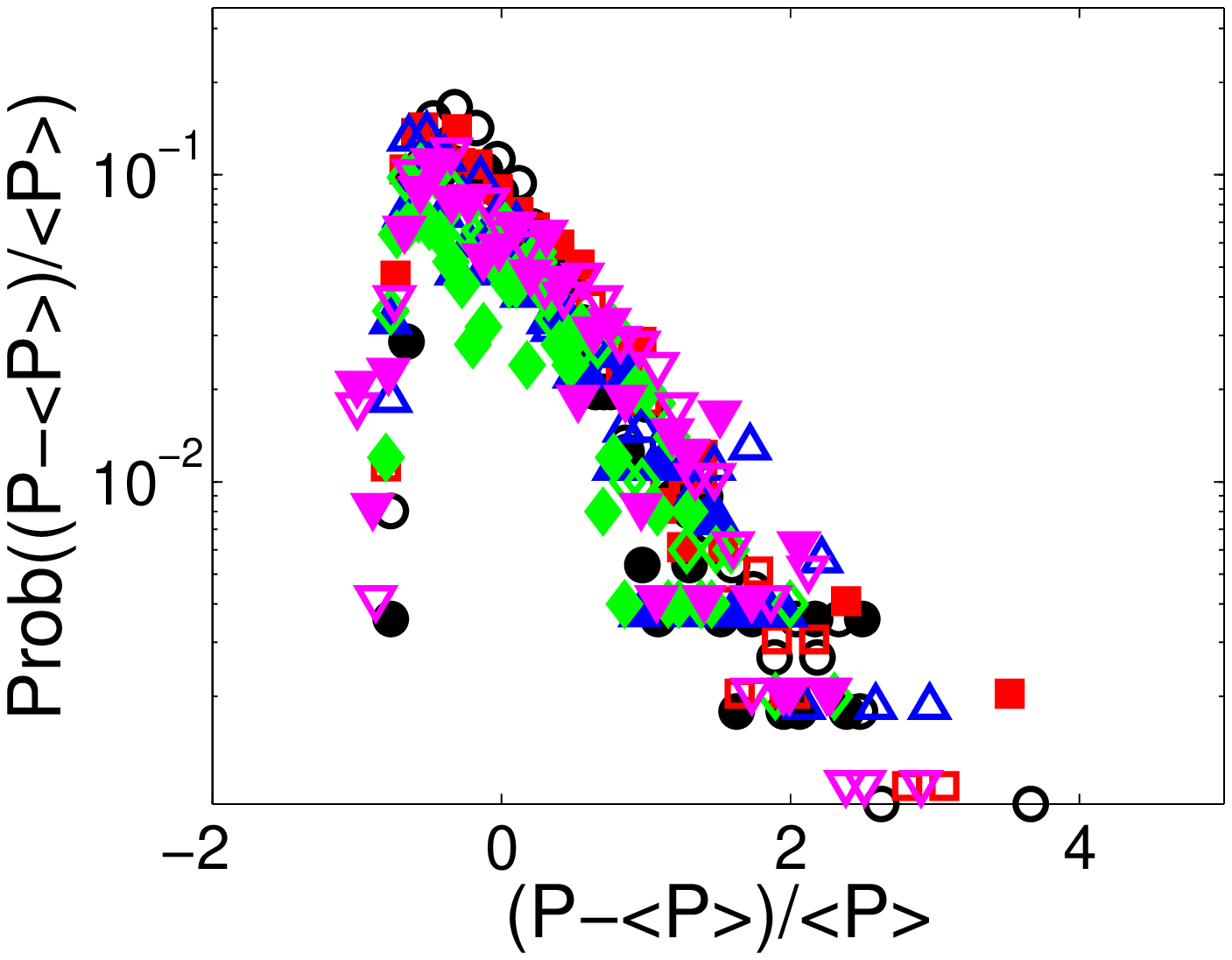}
\includegraphics[width=3.375 in]{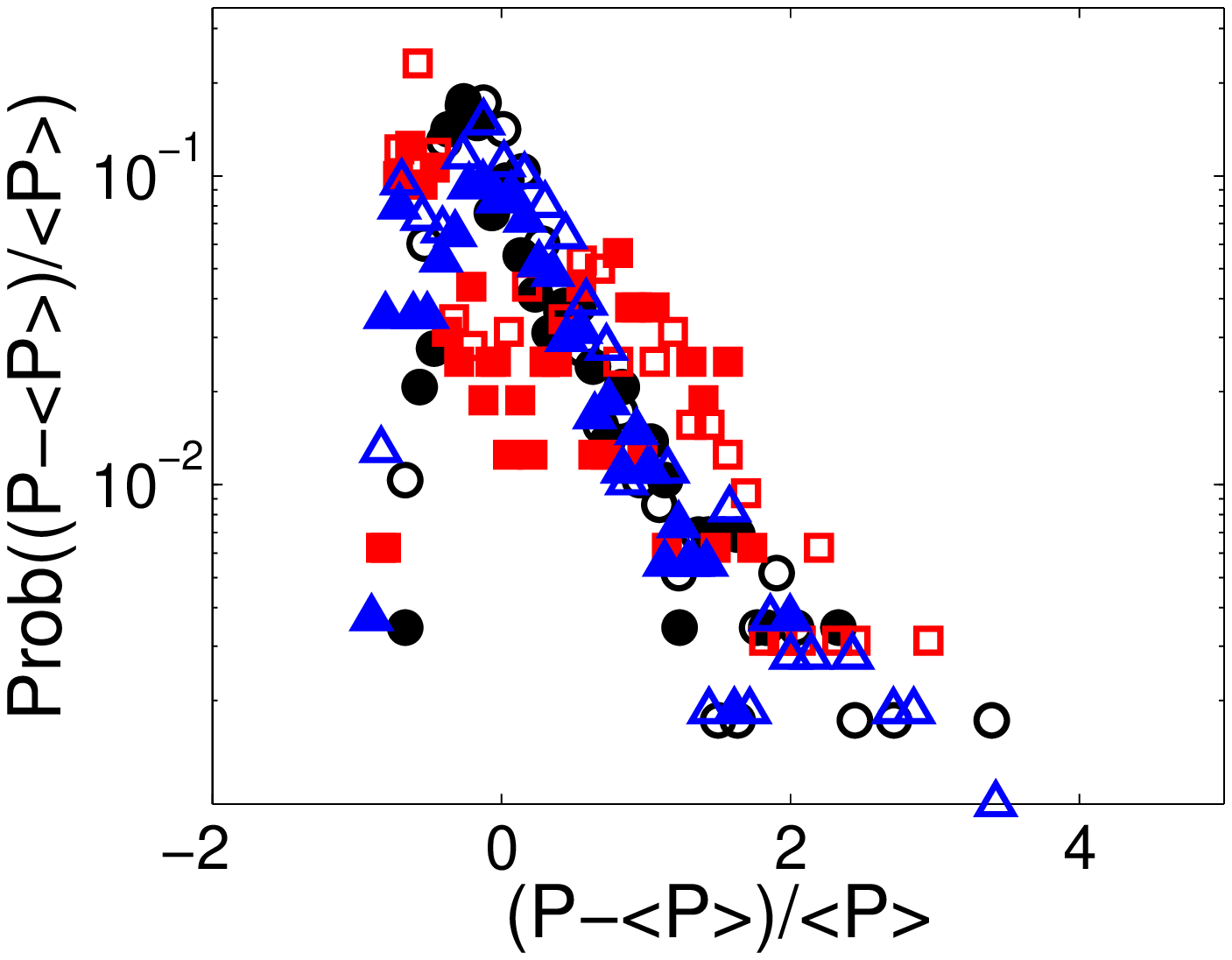}
\caption[Force fluctuations about mean profile]{Probability
distributions for the difference in pressure from the mean value at
the center ({\bf solid}) and sides ({\bf open}) of the silo for soft ({\bf
top}) particles in an unloaded silo ($\bullet$) and with overloads of $20$ g
($\blacksquare$) and $80$ g ($\blacktriangle$).  We also show distributions for
hard ({\bf bottom}) particles with overloads of $1837$ g ($\bullet$), $2267$ g
($\blacksquare$) and $3402$ g ($\blacktriangle$).  The curves differ
for low pressures due to the differences in sensitivity of the hard
and soft photoelastic polymers, but generally there is a great deal of overlap
and we cannot distinguish the sides from the center.}
\label{fluctuations}
\end{figure}

\subsection{Horizontal Variations and Fluctuations}

 It is also interesting to investigate various other aspects of
 the force profiles which differ from the usual Janssen picture.  In
 particular, we consider horizontal variability of the forces, and
 also force fluctuations.  In this regard, we note model studies by
 Pitman which exhibited horizontal dependence on fluctuations and
 stress within granular silos\cite{Pitman98} due to inhomogeneous
 mobilization of friction.

As shown in Fig.~\ref{fluctuations}, we measured the local deviation
from the mean pressure at a given height for both the center and sides
of the silo for all overloads for both hard and soft particles.  We
did not see significant horizontal variation in the distributions, and
for the soft particles the distributions were nearly identical for all
cases and very reminiscent of force distributions observed for other
granular systems \cite{Behringer05}.  Forces below the mean fall off
quickly, while forces larger than the mean tail off slowly, and
roughly follow an exponential.

\begin{figure} \centering
\includegraphics[width=3.375in]{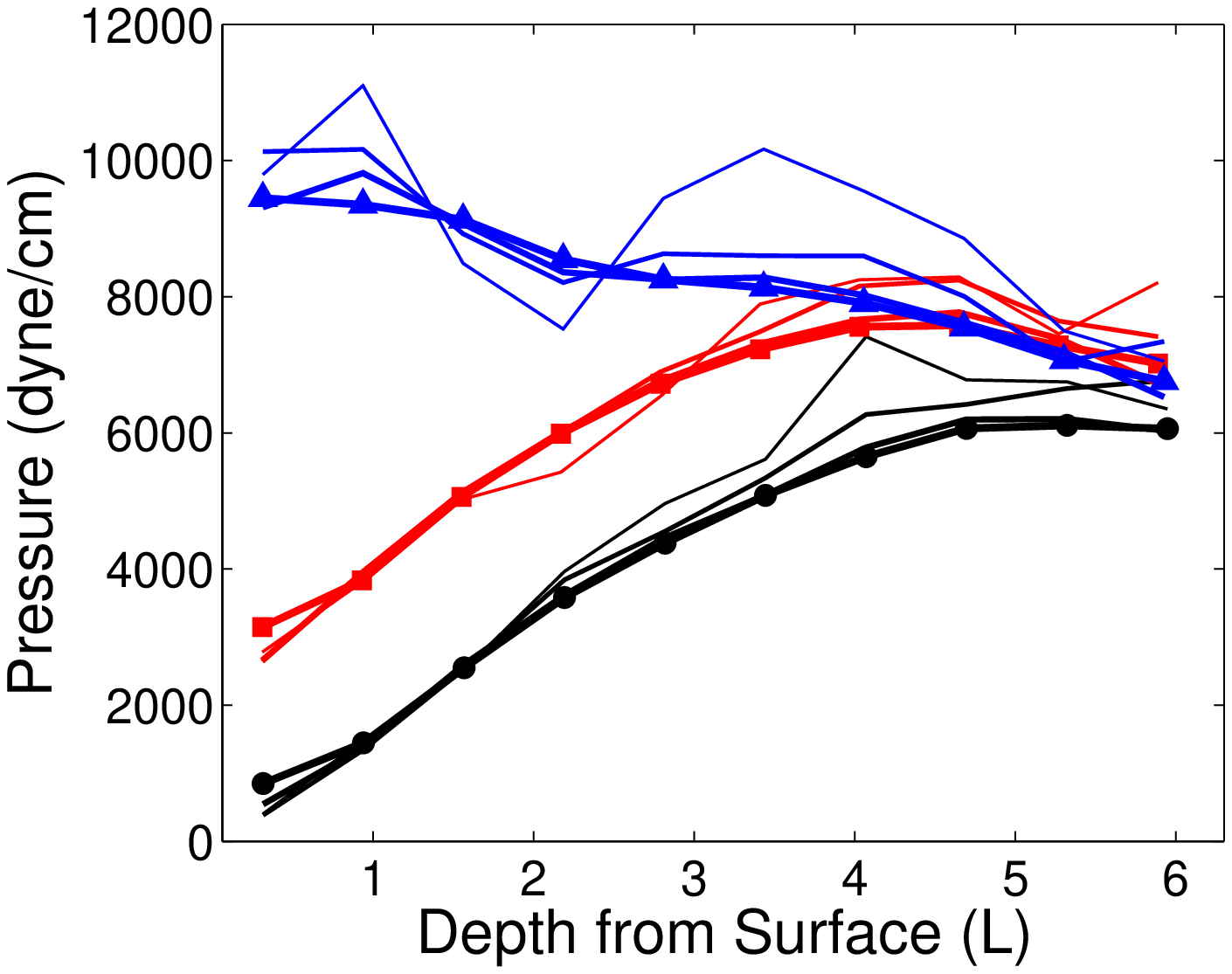}
\includegraphics[width=3.375in]{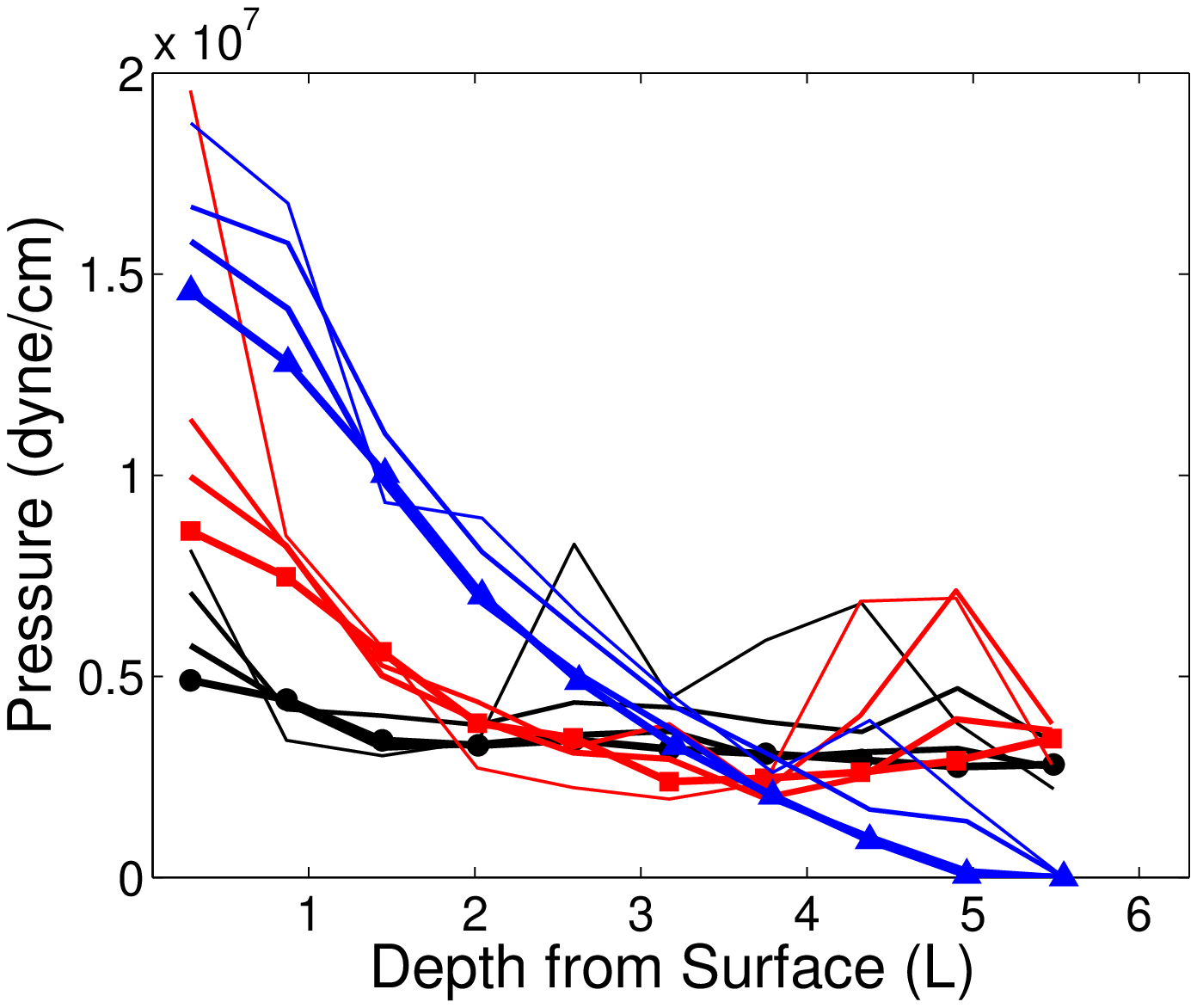}
\caption[Convergence of pressure profiles with averaging lengthscale]{Pressure
profiles on axis converge as the length $R$ over which averaging is
conducted varies, as indicated by lines of increasing thickness for $R =
1$, $5$, $10$ and $15$ small particle diameters for both soft particles
({\bf top}) in an unloaded silo
($\bullet$) and with overloads of $20$ g ($\blacksquare$) and $80$ g
($\blacktriangle$) and for hard particles ({\bf bottom}) with
overloads of $1837$ g ($\bullet$), $2267$ g ($\blacksquare$) and $3402$ g
($\blacktriangle$).}
\label{correlationplot} \end{figure}
Recent simulations for vertical granular matter with periodic boundary
conditions (in the horizontal directions) have indicated that the
averaging length-scale sufficient to determine the mean pressure is
possibly as small as a grain diameter \cite {Goldhirsch05b}.  Though
our geometry has side walls, it is still worthwhile to determine the
length over which we must average for the pressure profile to
converge.  To do this, we determine a grid of points within the silo
and average the pressure within a tunable radius to assign a pressure
to that point.  In Fig.~\ref{correlationplot}, we plot the pressure in
the middle of the silo (on axis) with depth as determined by averaging
over regions with radii of $1$, $5$, $10$ and $15$ small particle
diameters.  For the two larger averaging regions, the profiles
converge to those observed using horizontal binning by depth in
Fig.~\ref{newsoftpressure}.  This convergence indicates a correlation
length-scale between $10$ and $15$ particle diameters ---
approximately the same as the Janssen screening length-scale obtained
numerically.  This is interesting since in this instance, the
macroscopic scale of the Janssen parameter $\lambda$ is comparable to
the averaging scale, even though the Janssen model does not include
particle size.  As can be seen for the lower image in
Fig.~\ref{correlationplot}, the the effect of averaging length is
roughly the same regardless of particle stiffness.

An additional aspect of interest is the geometric nature of the
contacts.  Relevant quantities include the packing fraction, $\phi$,
the number of contacts per particle, $Z$, and clustering coefficient,
defined below.

With our second preparation procedure, we are able to obtain to obtain
all of these quantities.  We do this by capturing both polarized images
showing the force network within the photoelastic disks and
unpolarized images showing the locations of those particles.  Particle
centers can be identified using a convolution with separate kernels
for the two particle sizes.  We then use a numerical algorithm to
reject overlapping particles and identify neighbor particles in
contact with each particle based upon the geometric consideration that
the distance between two particles must be equal to the sum of their
radii.  This allows us to determine a local packing fraction
throughout the silo.

\begin{figure}
\centering
\includegraphics[width=3.3in]{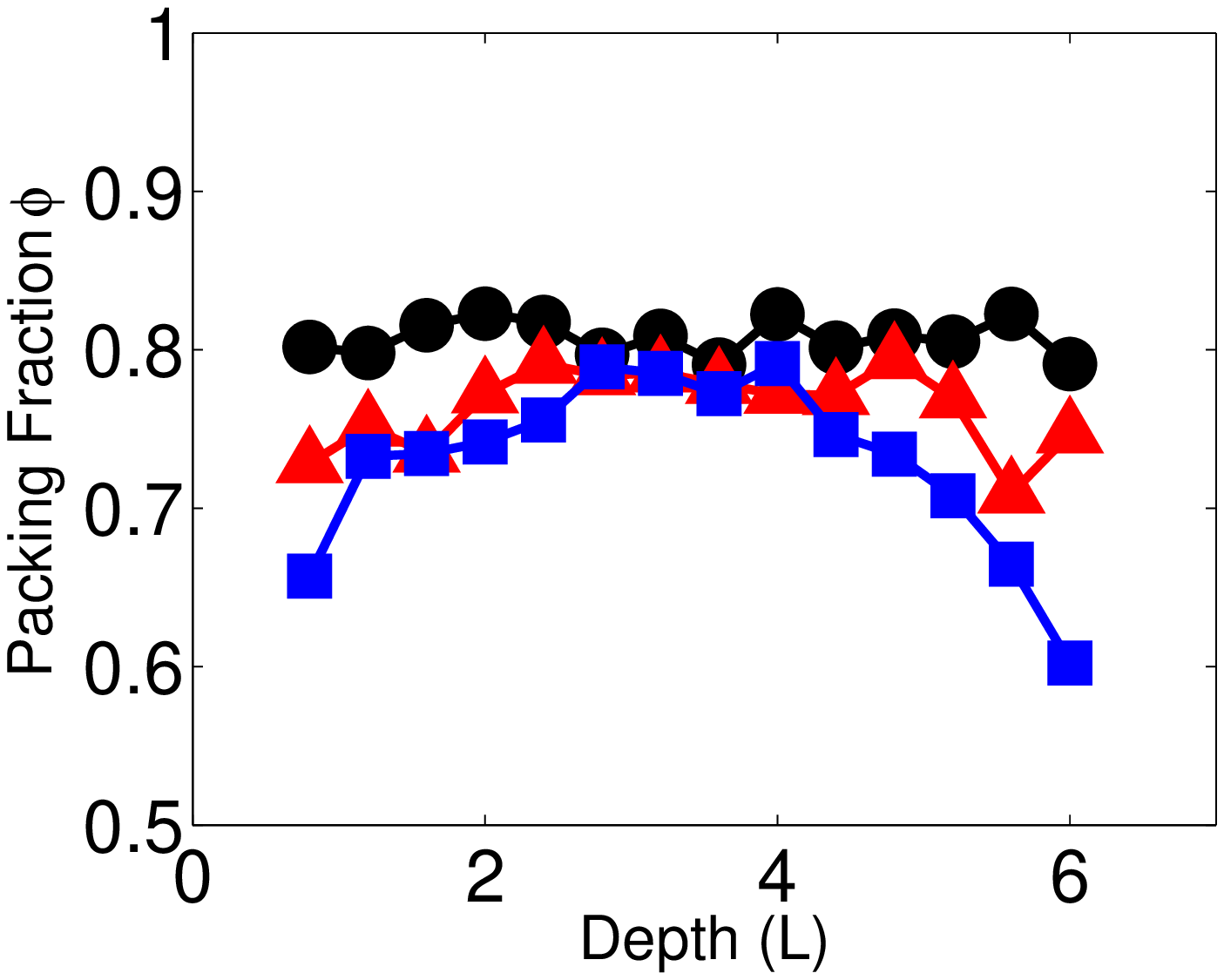}
\includegraphics[width=3.3in]{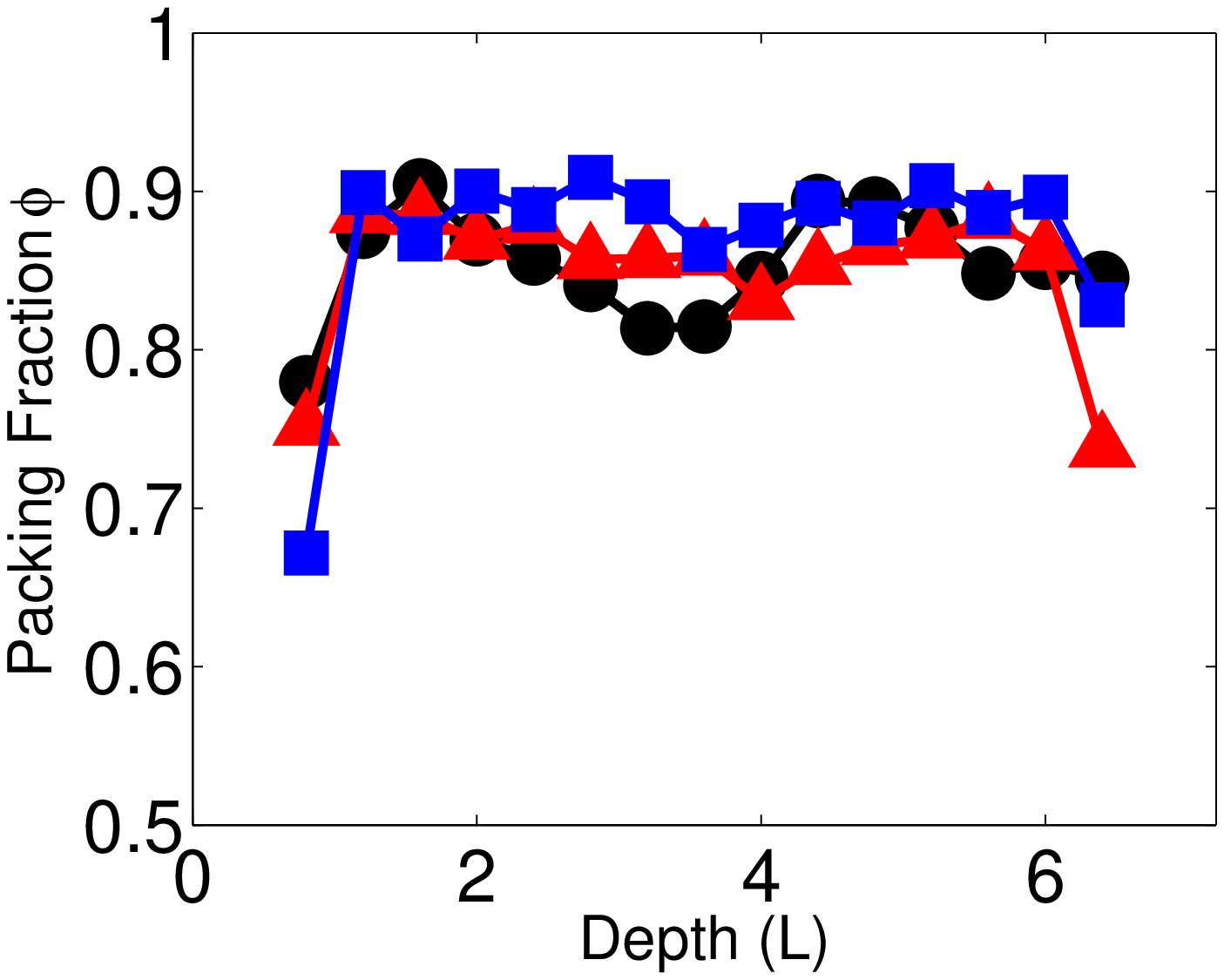}
\caption[Packing fraction with depth]{The average packing fraction for
soft ({\bf left}) and hard ({\bf right}) particles varies slightly
depending upon overload.  While it is roughly constant with depth for
an unloaded silo ($\bullet$) it varies for overloads of $20$ g
($\blacksquare$) and $80$ g ($\blacktriangle$).}
\label{phiplot}
\end{figure}
Since previous studies have shown that the packing fraction of a pile
relates to the redirection parameter, we investigate depth-dependent
variations of the packing fraction \cite{Clement03}.  Due to the dilation of the material, we find that the packing fraction is slightly lower for loaded silos for the depths available in the unloaded case.  We do not find systematic correlation between
the packing fraction and the screening length-scale.

The particle locations and contacts comprise a contact network for
each pile.  Analyzing these contact networks allows the determination
of graph and network properties such as the average number of contacts
per particle, $Z$.  We can also determine the `clustering
coefficient', which gives a measure of how tightly interconnected is
the network \cite{Watts98}.  For each particle we consider the $n$
neighboring particles with which that particle is in
contact.  There are $n-1$ possible neighbor-neighbor contacts for
round particles in two dimensions, so we define the clustering coefficient as the ratio
of the $m$ actual contacts present between those neighbors to the
$n-1$ possible contacts, $C \equiv \frac{m}{n-1}$.

For each contact network, we can determine a subset of contacts for
which the particle-scale pressure is greater than the mean for
the whole silo.  This subset of contacts can be thought of as roughly
comprising a force network, for which the same graph and network
properties as the contact network can also be calculated
\cite{Wambaugh06a}.

\begin{figure}
\centering \includegraphics[width=3.375in]{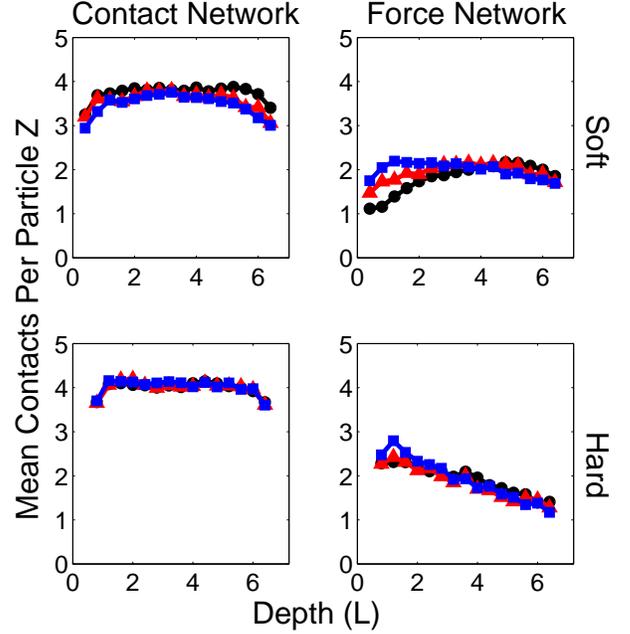}
\caption[Average contact number with depth for particles in contact
and force networks]{The average number of contacts per particle is
roughly constant for the contact network ({\bf left}) but varies at
the top of the pile for the force network ({\bf right}).  This is true
for both soft particles ({\bf above}) in an unloaded silo ($\bullet$)
and with overloads of $20$ g ($\blacksquare$) and $80$ g
($\blacktriangle$) as a well as for hard particles ({\bf below}) with
overloads of $1837$ g ($\bullet$), $2267$ g ($\blacksquare$) and
$3402$ g ($\blacktriangle$).}
\label{kplots}
\end{figure}
We first examine the dependence of the mean number of contacts per
particle with depth for both the overall contact network and the force
network.  As shown in Fig.~\ref{kplots}, we find that for the contact
network, the number of contacts is nearly independent of
overload and constant with depth at a value close to $k = 4$, the
typically observed coordination number for a disordered granular
material, although slightly lower due to the presence of the walls
\cite{Roux89}.  The average number of contacts in the force network is
slightly less than to two, which is not surprising since many particles
in a chain have only two contacts and, in our relative small silo, many particles
make contact with the walls.

Significantly, for the force network within both the hard and soft
particles, the average number of contacts at the top of the pile
varies depending upon the overload, before converging with depth,
unlike the contact network which does not vary with overload.  This
indicates that the force network is quantitatively different,
depending upon the size of the applied overload.  For the
deep-penetrating overload on the soft particles, the mean number of
contacts at the top is closest to the saturation value, perhaps
indicating that the structure of the force network is made visible by
the additional pressure, but is not disturbed as in the case of larger
overloads.

\begin{figure}
\centering
\includegraphics[width=3.375in]{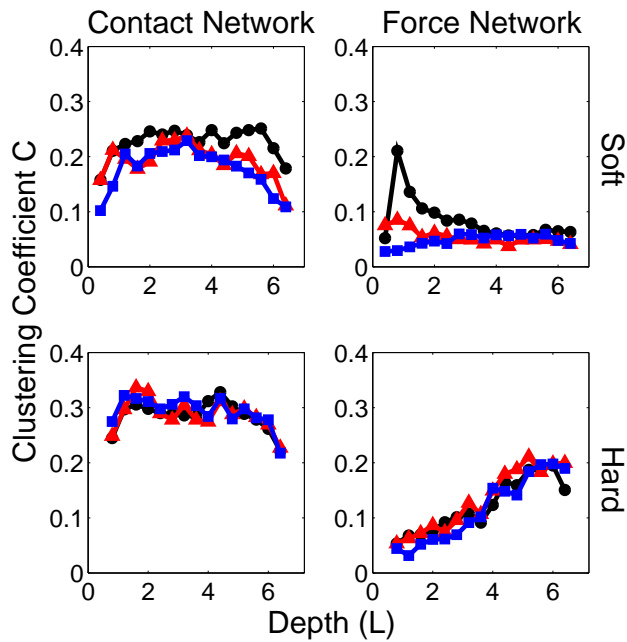}
\caption[Average clustering coefficient with depth for contact and
force networks]{For the contact network ({\bf left}) in soft ({\bf
above}) and hard ({\bf below}) particles the average clustering
coefficient is roughly the same for an unloaded silo ($\bullet$) and
overloads of $20$ g ($\blacksquare$) and $80$ g ($\blacktriangle$) for
soft particles and $1837$ g ($\bullet$), $2267$ g ($\blacksquare$) and
$3402$ g ($\blacktriangle$) for hard particles.  For the force network
({\bf right}) within soft particles, the clustering coefficient varies
depending upon overload.  For hard particles the force network is
unchanged by overload.}
\label{Cplots}
\end{figure}
For the clustering coefficient, Fig.~\ref{Cplots} shows the contact
network to be better connected than the force network.  As with the
number of contacts, the clustering coefficient is roughly constant for
the contact network but varies with depth for the force network.
For hard particles, the
overloads do not seem to alter the clustering coefficient of the
contact or force network.

The differences between the contact and force networks may indicate an
explanation for the deep propagation of small overloads.  If the
material within a silo can be characterized as fragile matter, then we
expect that it will have the ability to support forces along the force
chains that are present, but that the force network will have to
rearrange to support forces along normal directions \cite{Claudin98}
or to carry large loads.  This is borne out for our softer particles
which are more photoelastically sensitive.  We observe
deep-penetrating response for small loads for which there is only
slight deformation of the network, while larger loads rearrange the
force network, resulting in Janssen screening.

\section{Conclusion}

For elastic particles in granular silos, we observe a non-linear
response as we vary the size of overloads placed upon the material at
the top of the silo.  For sufficiently small overloads, the
pre-existing contact network conveys additional pressure deep into the
pile, contrary to the classical Janssen description.  For larger
overloads we find that the force network rearranges, creating
frictional arches braced against the walls that act to screen the
depths of the silo from the additional pressure.  Though the Janssen
model predicts monotonic saturation, we find that the pressure
initially increases with depth before decaying --- the giant
overshoot phenomenon \cite{Clement03}.

By numerically evaluating the Janssen length-scale we find that
it is not constant, meaning that the simple integration typically used
to derive pressure profile predictions from force balance is
inadequate in our experiments, and therefore quite likely in general.
We find that there is a different Janssen length-scale at the top of
the pile, possibly corresponding to changes in mobilization resulting
from placing the overload onto the material.

We find that the pressure profile is sensitive to preparation, and
that this sensitivity depends on both loading and particle stiffness.
The greatest dependence on history occurs when the magnitudes of
gravitational loading and the applied loading are commensurate.  This
situation occurs for our softer particles, for which we apply a
maximum load that corresponds to roughly half the weight of the
particles.  For our harder particles, which have weight comparable to
our softer particles, all applied loads significantly exceed the
particle weight.  Such loads lead to particle deformations comparable
to the soft case, but the observed non-linear behavior of the profiles
is substantially less.

We find that distributions for the particle-scale pressure are similar
to those observed for local force measures in other granular systems.
Specifically, there is a roughly exponential decay of the
distributions at large pressures.  There is a peak in the
distributions near the mean, and a rapid fall off below the mean.  In
general, we find that force networks exhibit fluctuations with a
length-scale comparable to the width of the silo.

\section*{Acknowledgements}
We appreciate work by Evelyne Kolb in the initial states of this research.
Additionally, we thank Annie Thebprasith, Trush Majmudar, Brian Tighe, and Karen
Daniels for help developing the instrumentation and many useful
discussions.  This research was funded by National Science Foundation
grants DMR-0137119, DMR-0555431, DMS-0204677, and NASA grant
NNC04GB08G.

\bibliography{granular}

\begin{thebibliography}{34}
\expandafter\ifx\csname natexlab\endcsname\relax\def\natexlab#1{#1}\fi
\expandafter\ifx\csname bibnamefont\endcsname\relax
  \def\bibnamefont#1{#1}\fi
\expandafter\ifx\csname bibfnamefont\endcsname\relax
  \def\bibfnamefont#1{#1}\fi
\expandafter\ifx\csname citenamefont\endcsname\relax
  \def\citenamefont#1{#1}\fi
\expandafter\ifx\csname url\endcsname\relax
  \def\url#1{\texttt{#1}}\fi
\expandafter\ifx\csname urlprefix\endcsname\relax\def\urlprefix{URL }\fi
\providecommand{\bibinfo}[2]{#2}
\providecommand{\eprint}[2][]{\url{#2}}

\bibitem[{\citenamefont{Vanel et~al.}(1999)\citenamefont{Vanel, Howell, Clark,
  Behringer, and Cl\'{e}ment}}]{Clement99a}
\bibinfo{author}{\bibfnamefont{L.}~\bibnamefont{Vanel}},
  \bibinfo{author}{\bibfnamefont{D.}~\bibnamefont{Howell}},
  \bibinfo{author}{\bibfnamefont{D.}~\bibnamefont{Clark}},
  \bibinfo{author}{\bibfnamefont{R.~P.} \bibnamefont{Behringer}},
  \bibnamefont{and}
  \bibinfo{author}{\bibfnamefont{E.}~\bibnamefont{Cl\'{e}ment}},
  \bibinfo{journal}{Physical Review E} \textbf{\bibinfo{volume}{60}},
  \bibinfo{pages}{R5040} (\bibinfo{year}{1999}).

\bibitem[{\citenamefont{Vanel et~al.}(2000)\citenamefont{Vanel, Claudin,
  Bouchaud, Cates, Cl\'{e}ment, and Wittmer}}]{Clement00}
\bibinfo{author}{\bibfnamefont{L.}~\bibnamefont{Vanel}},
  \bibinfo{author}{\bibfnamefont{P.}~\bibnamefont{Claudin}},
  \bibinfo{author}{\bibfnamefont{J.-P.} \bibnamefont{Bouchaud}},
  \bibinfo{author}{\bibfnamefont{M.~E.} \bibnamefont{Cates}},
  \bibinfo{author}{\bibfnamefont{E.}~\bibnamefont{Cl\'{e}ment}},
  \bibnamefont{and} \bibinfo{author}{\bibfnamefont{J.~P.}
  \bibnamefont{Wittmer}}, \bibinfo{journal}{Physical Review Letters}
  \textbf{\bibinfo{volume}{84}}, \bibinfo{pages}{1439} (\bibinfo{year}{2000}).

\bibitem[{\citenamefont{Geng et~al.}(2003)\citenamefont{Geng, Reydellet,
  Cl\'{e}ment, and Behringer}}]{Behringer03}
\bibinfo{author}{\bibfnamefont{J.}~\bibnamefont{Geng}},
  \bibinfo{author}{\bibfnamefont{G.}~\bibnamefont{Reydellet}},
  \bibinfo{author}{\bibfnamefont{E.}~\bibnamefont{Cl\'{e}ment}},
  \bibnamefont{and} \bibinfo{author}{\bibfnamefont{R.~P.}
  \bibnamefont{Behringer}}, \bibinfo{journal}{Physica D}
  \textbf{\bibinfo{volume}{182}}, \bibinfo{pages}{274} (\bibinfo{year}{2003}).

\bibitem[{\citenamefont{Mueth et~al.}(1998)\citenamefont{Mueth, Jaeger, and
  Nagel}}]{Mueth98}
\bibinfo{author}{\bibfnamefont{D.~M.} \bibnamefont{Mueth}},
  \bibinfo{author}{\bibfnamefont{H.~M.} \bibnamefont{Jaeger}},
  \bibnamefont{and} \bibinfo{author}{\bibfnamefont{S.~R.} \bibnamefont{Nagel}},
  \bibinfo{journal}{Physical Review E} \textbf{\bibinfo{volume}{57}},
  \bibinfo{pages}{3164} (\bibinfo{year}{1998}).

\bibitem[{\citenamefont{Janssen}(1895)}]{Janssen}
\bibinfo{author}{\bibfnamefont{H.~A.} \bibnamefont{Janssen}},
  \bibinfo{journal}{Zeitschr. d. Vereines deutscher Ingenieure}
  \textbf{\bibinfo{volume}{39}}, \bibinfo{pages}{1045} (\bibinfo{year}{1895}).

\bibitem[{\citenamefont{Sperl}(2006)}]{Sperl05}
\bibinfo{author}{\bibfnamefont{M.}~\bibnamefont{Sperl}},
  \bibinfo{journal}{Granular Matter} \textbf{\bibinfo{volume}{8}},
  \bibinfo{pages}{59} (\bibinfo{year}{2006}).

\bibitem[{\citenamefont{Nedderman}(1992)}]{Nedderman92}
\bibinfo{author}{\bibfnamefont{R.}~\bibnamefont{Nedderman}},
  \emph{\bibinfo{title}{Statics and Kinematics of Granular Materials}}
  (\bibinfo{publisher}{Cambridge University Press}, \bibinfo{year}{1992}).

\bibitem[{\citenamefont{Ovarlez et~al.}(2003)\citenamefont{Ovarlez, Fond, and
  Cl\'{e}ment}}]{Clement03}
\bibinfo{author}{\bibfnamefont{G.}~\bibnamefont{Ovarlez}},
  \bibinfo{author}{\bibfnamefont{C.}~\bibnamefont{Fond}}, \bibnamefont{and}
  \bibinfo{author}{\bibfnamefont{E.}~\bibnamefont{Cl\'{e}ment}},
  \bibinfo{journal}{Physical Review E} \textbf{\bibinfo{volume}{67}},
  \bibinfo{pages}{060302(R)} (\bibinfo{year}{2003}).

\bibitem[{\citenamefont{Bertho et~al.}(2003)\citenamefont{Bertho,
  Giorgiutti-Dauphin\'{e}, and Hulin}}]{Hulin03}
\bibinfo{author}{\bibfnamefont{Y.}~\bibnamefont{Bertho}},
  \bibinfo{author}{\bibfnamefont{F.}~\bibnamefont{Giorgiutti-Dauphin\'{e}}},
  \bibnamefont{and} \bibinfo{author}{\bibfnamefont{J.-P.} \bibnamefont{Hulin}},
  \bibinfo{journal}{Physical Review Letters} \textbf{\bibinfo{volume}{90}},
  \bibinfo{pages}{144301} (\bibinfo{year}{2003}).

\bibitem[{\citenamefont{Vanel and Cl\'{e}ment}(1999)}]{Clement99b}
\bibinfo{author}{\bibfnamefont{L.}~\bibnamefont{Vanel}} \bibnamefont{and}
  \bibinfo{author}{\bibfnamefont{E.}~\bibnamefont{Cl\'{e}ment}},
  \bibinfo{journal}{European Physical Journal B} \textbf{\bibinfo{volume}{11}},
  \bibinfo{pages}{525} (\bibinfo{year}{1999}).

\bibitem[{\citenamefont{Bertho et~al.}(2004)\citenamefont{Bertho, Brunet,
  Giorgiutti-Dauphin\'{e}, and Hulin}}]{Hulin04}
\bibinfo{author}{\bibfnamefont{Y.}~\bibnamefont{Bertho}},
  \bibinfo{author}{\bibfnamefont{T.}~\bibnamefont{Brunet}},
  \bibinfo{author}{\bibfnamefont{F.}~\bibnamefont{Giorgiutti-Dauphin\'{e}}},
  \bibnamefont{and} \bibinfo{author}{\bibfnamefont{J.-P.} \bibnamefont{Hulin}},
  \bibinfo{journal}{Europhysics Letters} \textbf{\bibinfo{volume}{67}},
  \bibinfo{pages}{955} (\bibinfo{year}{2004}).

\bibitem[{\citenamefont{Cl\'{e}ment et~al.}(1997)\citenamefont{Cl\'{e}ment,
  Serero, Lanuza, Rajchenbach, and Duran}}]{Clement97}
\bibinfo{author}{\bibfnamefont{E.}~\bibnamefont{Cl\'{e}ment}},
  \bibinfo{author}{\bibfnamefont{Y.}~\bibnamefont{Serero}},
  \bibinfo{author}{\bibfnamefont{J.}~\bibnamefont{Lanuza}},
  \bibinfo{author}{\bibfnamefont{J.}~\bibnamefont{Rajchenbach}},
  \bibnamefont{and} \bibinfo{author}{\bibfnamefont{J.}~\bibnamefont{Duran}}, in
  \emph{\bibinfo{booktitle}{Powders and Grains}}, edited by
  \bibinfo{editor}{\bibfnamefont{R.~P.} \bibnamefont{Behringer}}
  \bibnamefont{and} \bibinfo{editor}{\bibfnamefont{J.~T.}
  \bibnamefont{Jenkins}} (\bibinfo{year}{1997}), pp. \bibinfo{pages}{349--351}.

\bibitem[{\citenamefont{Landry et~al.}(2003)\citenamefont{Landry, Grest,
  Silbert, and Plimpton}}]{Landry03a}
\bibinfo{author}{\bibfnamefont{J.~W.} \bibnamefont{Landry}},
  \bibinfo{author}{\bibfnamefont{G.~S.} \bibnamefont{Grest}},
  \bibinfo{author}{\bibfnamefont{L.~E.} \bibnamefont{Silbert}},
  \bibnamefont{and} \bibinfo{author}{\bibfnamefont{S.~J.}
  \bibnamefont{Plimpton}}, \bibinfo{journal}{Physical Review E}
  \textbf{\bibinfo{volume}{67}}, \bibinfo{pages}{041303}
  (\bibinfo{year}{2003}).

\bibitem[{\citenamefont{Peralta-Fabi et~al.}(1997)\citenamefont{Peralta-Fabi,
  M\'{a}laga, and Rechtman}}]{Peralta-Fabi97}
\bibinfo{author}{\bibfnamefont{R.}~\bibnamefont{Peralta-Fabi}},
  \bibinfo{author}{\bibfnamefont{C.}~\bibnamefont{M\'{a}laga}},
  \bibnamefont{and} \bibinfo{author}{\bibfnamefont{R.}~\bibnamefont{Rechtman}},
  in \emph{\bibinfo{booktitle}{Powders and Grains}}, edited by
  \bibinfo{editor}{\bibfnamefont{R.~P.} \bibnamefont{Behringer}}
  \bibnamefont{and} \bibinfo{editor}{\bibfnamefont{J.~T.}
  \bibnamefont{Jenkins}} (\bibinfo{year}{1997}), pp. \bibinfo{pages}{227--230}.

\bibitem[{\citenamefont{Boutreux et~al.}(1997)\citenamefont{Boutreux,
  Rapha\"{e}l, and de~Gennes}}]{deGennes97}
\bibinfo{author}{\bibfnamefont{T.}~\bibnamefont{Boutreux}},
  \bibinfo{author}{\bibfnamefont{E.}~\bibnamefont{Rapha\"{e}l}},
  \bibnamefont{and} \bibinfo{author}{\bibfnamefont{P.~G.}
  \bibnamefont{de~Gennes}}, \bibinfo{journal}{Physical Review E.}
  \textbf{\bibinfo{volume}{55}}, \bibinfo{pages}{5759} (\bibinfo{year}{1997}).

\bibitem[{\citenamefont{Pitman}(1998)}]{Pitman98}
\bibinfo{author}{\bibfnamefont{E.~B.} \bibnamefont{Pitman}},
  \bibinfo{journal}{Physical Review E} \textbf{\bibinfo{volume}{57}},
  \bibinfo{pages}{3170} (\bibinfo{year}{1998}).

\bibitem[{\citenamefont{Landry and Grest}(2004)}]{Landry03c}
\bibinfo{author}{\bibfnamefont{J.~W.} \bibnamefont{Landry}} \bibnamefont{and}
  \bibinfo{author}{\bibfnamefont{G.~S.} \bibnamefont{Grest}},
  \bibinfo{journal}{Physical Review E} \textbf{\bibinfo{volume}{69}},
  \bibinfo{pages}{031303} (\bibinfo{year}{2004}).

\bibitem[{\citenamefont{Atewologun and Riskowski}(1991)}]{Atewologun91}
\bibinfo{author}{\bibfnamefont{A.~O.} \bibnamefont{Atewologun}}
  \bibnamefont{and}
  \bibinfo{author}{\bibfnamefont{G.}~\bibnamefont{Riskowski}},
  \bibinfo{journal}{Transactions of the ASAE} \textbf{\bibinfo{volume}{34}},
  \bibinfo{pages}{2193} (\bibinfo{year}{1991}).

\bibitem[{\citenamefont{Rusinek}(2003)}]{Rusinek03}
\bibinfo{author}{\bibfnamefont{R.}~\bibnamefont{Rusinek}},
  \bibinfo{journal}{Research in Agricultural Engineering}
  \textbf{\bibinfo{volume}{49}}, \bibinfo{pages}{61} (\bibinfo{year}{2003}).

\bibitem[{\citenamefont{Lvin}(1970)}]{Lvin70}
\bibinfo{author}{\bibfnamefont{J.~B.} \bibnamefont{Lvin}},
  \bibinfo{journal}{Powder Technology} \textbf{\bibinfo{volume}{4}},
  \bibinfo{pages}{280} (\bibinfo{year}{1970}).

\bibitem[{\citenamefont{Ovarlez and Cl\'{e}ment}(2005)}]{Clement05}
\bibinfo{author}{\bibfnamefont{G.}~\bibnamefont{Ovarlez}} \bibnamefont{and}
  \bibinfo{author}{\bibfnamefont{E.}~\bibnamefont{Cl\'{e}ment}},
  \bibinfo{journal}{European Physical Journal E} \textbf{\bibinfo{volume}{16}},
  \bibinfo{pages}{421} (\bibinfo{year}{2005}).

\bibitem[{\citenamefont{Cates et~al.}(1998)\citenamefont{Cates, Wittmer,
  Bouchaud, and Claudin}}]{Claudin98}
\bibinfo{author}{\bibfnamefont{M.~E.} \bibnamefont{Cates}},
  \bibinfo{author}{\bibfnamefont{J.~P.} \bibnamefont{Wittmer}},
  \bibinfo{author}{\bibfnamefont{J.-P.} \bibnamefont{Bouchaud}},
  \bibnamefont{and} \bibinfo{author}{\bibfnamefont{P.}~\bibnamefont{Claudin}},
  \bibinfo{journal}{Physical Review Letters} \textbf{\bibinfo{volume}{81}},
  \bibinfo{pages}{1841} (\bibinfo{year}{1998}).

\bibitem[{\citenamefont{Ovarlez et~al.}(2001)\citenamefont{Ovarlez, Kolb, and
  Cl\'{e}ment}}]{Clement01d}
\bibinfo{author}{\bibfnamefont{G.}~\bibnamefont{Ovarlez}},
  \bibinfo{author}{\bibfnamefont{E.}~\bibnamefont{Kolb}}, \bibnamefont{and}
  \bibinfo{author}{\bibfnamefont{E.}~\bibnamefont{Cl\'{e}ment}},
  \bibinfo{journal}{Physical Review E} \textbf{\bibinfo{volume}{64}},
  \bibinfo{pages}{060302(R)} (\bibinfo{year}{2001}).

\bibitem[{\citenamefont{Coppersmith et~al.}(1996)\citenamefont{Coppersmith,
  Liu, Majumdar, Narayan, and Witten}}]{Coppersmith96}
\bibinfo{author}{\bibfnamefont{S.~N.} \bibnamefont{Coppersmith}},
  \bibinfo{author}{\bibfnamefont{C.-h.} \bibnamefont{Liu}},
  \bibinfo{author}{\bibfnamefont{S.}~\bibnamefont{Majumdar}},
  \bibinfo{author}{\bibfnamefont{O.}~\bibnamefont{Narayan}}, \bibnamefont{and}
  \bibinfo{author}{\bibfnamefont{T.~A.} \bibnamefont{Witten}},
  \bibinfo{journal}{Physical Review E} \textbf{\bibinfo{volume}{53}},
  \bibinfo{pages}{4673} (\bibinfo{year}{1996}).

\bibitem[{\citenamefont{Socolar}(1997)}]{Socolar97}
\bibinfo{author}{\bibfnamefont{J.~E.~S.} \bibnamefont{Socolar}},
  \bibinfo{journal}{Physical Review E} \textbf{\bibinfo{volume}{57}},
  \bibinfo{pages}{3204} (\bibinfo{year}{1997}).

\bibitem[{\citenamefont{Baxter}(1997)}]{Baxter97}
\bibinfo{author}{\bibfnamefont{G.~W.} \bibnamefont{Baxter}}, in
  \emph{\bibinfo{booktitle}{Powders and Grains}}, edited by
  \bibinfo{editor}{\bibfnamefont{R.~P.} \bibnamefont{Behringer}}
  \bibnamefont{and} \bibinfo{editor}{\bibfnamefont{J.~T.}
  \bibnamefont{Jenkins}} (\bibinfo{year}{1997}), pp. \bibinfo{pages}{345--348}.

\bibitem[{\citenamefont{Landry et~al.}(2004)\citenamefont{Landry, Grest, and
  Plimpton}}]{Landry03b}
\bibinfo{author}{\bibfnamefont{J.~W.} \bibnamefont{Landry}},
  \bibinfo{author}{\bibfnamefont{G.~S.} \bibnamefont{Grest}}, \bibnamefont{and}
  \bibinfo{author}{\bibfnamefont{S.~J.} \bibnamefont{Plimpton}},
  \bibinfo{journal}{Powder Technology} \textbf{\bibinfo{volume}{139}},
  \bibinfo{pages}{233} (\bibinfo{year}{2004}).

\bibitem[{\citenamefont{Geng et~al.}(2001)\citenamefont{Geng, Howell, Longhi,
  Behringer, Reydellet, Vanel, Cl\'{e}ment, and Luding}}]{Clement01a}
\bibinfo{author}{\bibfnamefont{J.}~\bibnamefont{Geng}},
  \bibinfo{author}{\bibfnamefont{D.}~\bibnamefont{Howell}},
  \bibinfo{author}{\bibfnamefont{E.}~\bibnamefont{Longhi}},
  \bibinfo{author}{\bibfnamefont{R.~P.} \bibnamefont{Behringer}},
  \bibinfo{author}{\bibfnamefont{G.}~\bibnamefont{Reydellet}},
  \bibinfo{author}{\bibfnamefont{L.}~\bibnamefont{Vanel}},
  \bibinfo{author}{\bibfnamefont{E.}~\bibnamefont{Cl\'{e}ment}},
  \bibnamefont{and} \bibinfo{author}{\bibfnamefont{S.}~\bibnamefont{Luding}},
  \bibinfo{journal}{Physical Review Letters} \textbf{\bibinfo{volume}{87}},
  \bibinfo{pages}{035506} (\bibinfo{year}{2001}).

\bibitem[{\citenamefont{Efron and Tibshirani}(1991)}]{Efron91}
\bibinfo{author}{\bibfnamefont{B.}~\bibnamefont{Efron}} \bibnamefont{and}
  \bibinfo{author}{\bibfnamefont{R.}~\bibnamefont{Tibshirani}},
  \bibinfo{journal}{Science} \textbf{\bibinfo{volume}{253}},
  \bibinfo{pages}{390} (\bibinfo{year}{1991}).

\bibitem[{\citenamefont{Majmudar and Behringer}(2005)}]{Behringer05}
\bibinfo{author}{\bibfnamefont{T.~S.} \bibnamefont{Majmudar}} \bibnamefont{and}
  \bibinfo{author}{\bibfnamefont{R.~P.} \bibnamefont{Behringer}},
  \bibinfo{journal}{Nature} \textbf{\bibinfo{volume}{435}},
  \bibinfo{pages}{1079} (\bibinfo{year}{2005}).

\bibitem[{\citenamefont{Goldenberg et~al.}(2006)\citenamefont{Goldenberg,
  Atman, Claudin, Combe, and Goldhirsch}}]{Goldhirsch05b}
\bibinfo{author}{\bibfnamefont{C.}~\bibnamefont{Goldenberg}},
  \bibinfo{author}{\bibfnamefont{A.~P.~F.} \bibnamefont{Atman}},
  \bibinfo{author}{\bibfnamefont{P.}~\bibnamefont{Claudin}},
  \bibinfo{author}{\bibfnamefont{G.}~\bibnamefont{Combe}}, \bibnamefont{and}
  \bibinfo{author}{\bibfnamefont{I.}~\bibnamefont{Goldhirsch}},
  \bibinfo{journal}{Physical Review Letters} \textbf{\bibinfo{volume}{96}}
  (\bibinfo{year}{2006}).

\bibitem[{\citenamefont{Watts and Strogatz}(1998)}]{Watts98}
\bibinfo{author}{\bibfnamefont{D.~J.} \bibnamefont{Watts}} \bibnamefont{and}
  \bibinfo{author}{\bibfnamefont{S.~H.} \bibnamefont{Strogatz}},
  \bibinfo{journal}{Nature} \textbf{\bibinfo{volume}{393}},
  \bibinfo{pages}{440} (\bibinfo{year}{1998}).

\bibitem[{\citenamefont{Wambaugh}(2006)}]{Wambaugh06a}
\bibinfo{author}{\bibfnamefont{J.~F.} \bibnamefont{Wambaugh}},
  \bibinfo{journal}{cond-mat/0603314}  (\bibinfo{year}{2006}).

\bibitem[{\citenamefont{Roux and Hansen}(1989)}]{Roux89}
\bibinfo{author}{\bibfnamefont{S.}~\bibnamefont{Roux}} \bibnamefont{and}
  \bibinfo{author}{\bibfnamefont{A.}~\bibnamefont{Hansen}}, in
  \emph{\bibinfo{booktitle}{Powders and Grains}}, edited by
  \bibinfo{editor}{\bibfnamefont{J.}~\bibnamefont{Biarez}} \bibnamefont{and}
  \bibinfo{editor}{\bibfnamefont{R.}~\bibnamefont{Gourv\'{e}s}}
  (\bibinfo{year}{1989}), pp. \bibinfo{pages}{249--254}.

\end{thebibliography}
\end{document}